\documentclass{article}

\usepackage{PRIMEarxiv}

\usepackage[utf8]{inputenc} 
\usepackage[T1]{fontenc}    
\usepackage{hyperref}       
\usepackage{url}            
\usepackage{booktabs}       
\usepackage{amsfonts}       
\usepackage{nicefrac}       
\usepackage{microtype}      
\usepackage{lipsum}
\usepackage{fancyhdr}       
\usepackage{graphicx}       

\usepackage{longtable}
\usepackage{tabularx}
\usepackage{multirow}
\usepackage{multicol}
\usepackage{tabularray}
\usepackage{array}
\usepackage{caption}

\usepackage{hyperref}
\usepackage{subfigure}

\title{Deep PCCT: Photon Counting Computed Tomography Deep Learning Applications Review}
\author{
    Ana Carolina Alves\textsuperscript{1,2},
    André Ferreira\textsuperscript{1,2,3},
    Gijs Luijten\textsuperscript{2,4},
    Jens Kleesiek\textsuperscript{2,5},\\
    \textbf{Behrus Puladi\textsuperscript{3,6}}, 
    \textbf{Jan Egger\textsuperscript{2,4,7}},
    \textbf{Victor Alves\textsuperscript{1}}
}

\begin{document}

\maketitle
\textsuperscript{1} Center Algoritmi / LASI, University of Minho, Braga, Portugal

\textsuperscript{2} Institute for Artificial Intelligence in Medicine, University Medicine Essen, Essen, Germany

\textsuperscript{3} Department of Oral and Maxillofacial Surgery, University Hospital RWTH Aachen, Aachen, Germany

\textsuperscript{4} Institute of Computer Graphics and Vision (ICG), Graz University of Technology, Graz, Austria

\textsuperscript{5} Department of Physics, TU Dortmund University, Dortmund, Germany

\textsuperscript{6} Institute of Medical Informatics, University Hospital RWTH Aachen, Aachen, Germany

\textsuperscript{7} Center for Virtual and Extended Reality in Medicine, University Medicine Essen, Essen, Germany

\paragraph{}

\begin{abstract}
Medical imaging faces challenges such as limited spatial resolution, interference from electronic noise and poor contrast-to-noise ratios. Photon Counting Computed Tomography (PCCT) has emerged as a solution, addressing these issues with its innovative technology. This review delves into the recent developments and applications of PCCT in pre-clinical research, emphasizing its potential to overcome traditional imaging limitations. For example PCCT has demonstrated remarkable efficacy in improving the detection of subtle abnormalities in breast, providing a level of detail previously unattainable.
Examining the current literature on PCCT, it presents a comprehensive analysis of the technology, highlighting the main features of scanners and their varied applications. In addition, it explores the integration of deep learning into PCCT, along with the study of radiomic features, presenting successful applications in data processing. While acknowledging these advances, it also discusses the existing challenges in this field, paving the way for future research and improvements in medical imaging technologies.
Despite the limited number of articles on this subject, due to the recent integration of PCCT at a clinical level, its potential benefits extend to various diagnostic applications.
\end{abstract}


\paragraph{}

\textbf{\emph{Abbreviations:}}
APOE, Apolipoprotein E; AUC, Area under the ROC Curve; CCC, Concordance Correlation Coefficient; CT, Computed Tomography; CTDIvol, Computed Tomography Dose Index Volume; CNN, Convolutional Neural Networks; CNR, Contrast-to-Noise Ratio; DL, Deep Learning;
DECT, Dual Energy Computed Tomography; d'adj, Adjusted Detectability Index; EICT, Energy Integrating Computed Tomography; EID, Energy Integrating Detector; FDA, U.S. Food and Drug Administration; f10, 10\% Frequency; fav, Average Frequency; FBP, Filtered Back Projection; fpeak, Frequency Peak; GNI, Global Noise Index; HCC, Hepato Cellular Carcinoma; ICC, Intraclass Correlation Coefficient; iMAR, Iterative Reduction of Metallic Artifacts; LOOCV, Leave-One-Out Cross-Validation; ML, Machine Learning; MLR, Multinomial Logistic Regression; MPR, Major Pathological Response; NSCLC, Non-Small Cell Lung Cancer; NPS, Noise Power Spectrum; PCA, Principal Component Analysis; PCCT, Photon Counting Computed Tomography; PCD, Photon Counting Detector; PSNR, Peak Signal-to-Noise Ratio; QIR, Quantum Iterative Reconstruction; ROI, Region of Interest; SRCC, Spearman Rank Correlation Coefficient; SSIM, Structure Similarity; SVM, Support Vector Machine; TTF, Transfer Time Function; VMI, Virtual Monoenergetic Imaging; VNC, Virtual Non-Contrast; VOI, Volume of Interest.

\section{Introduction}
\paragraph{}
Since the discovery of Hounsfield and the first CT scan, it has become an integral part of contemporary medicine. The quality of the medical images has improved along with the gradual improvement of  scanning techniques. However, these improvements are still insufficient to distinguish crucial differences between biological tissues and contrast agents due to lacking spatial resolution. Over the last decade, spectral CT has been developed as a new generation of CT technology, with the latest development being the PCCT.
The first clinically approved PCCT system, receiving FDA authorization in September 2021 marked a milestone in medical imaging.
PCCT is a computed tomography modality that uses a PCD to register the interactions of individual photons, allowing precise monitoring of the energy used in each interaction.
This enables the detector of a PCD to record and approximate energy spectrum of the individual registered photon. Therefore the PCCT is seen as a spectral or energy-resolved CT technique which counts the individual photons and categorizes them into energy bins based on the heights of the incoming electric pulses. This technique thus allows for filtering out the electronic noise otherwise indistinguishable from the original electronic signal generated by the incoming photon \cite{Review1_judith}, \cite{Bousse2023}, \cite{Review2_willemink}.
\subparagraph{}
Medical image processing and analysis encompass various tasks, such as image retrieval, creation, analysis and visualization. The field of medical image processing has expanded to incorporate computer vision, pattern recognition, image mining and ML in various applications. Deep learning has become a predominant methodology, contributing to the accuracy of subsequent steps. This has given rise to new possibilities for advances in medical image analysis \cite{Razzak2018}. Another approach is radiomics studies, which seek to extract information and identify clinically relevant features from radiological imaging data that can be difficult for the human eye to discern. To facilitate detection, ML and DL methods are being increasingly researched, with the aim of improving patient diagnosis, treatment options and overall results \cite{Wagner2021}. 
For example, radiomic features that may escape visual detection can be identified and analysed by sophisticated computer algorithms, providing valuable information for clinical decision-making, as in the study by Kahmann et al. which identified two distinct radiomic features, specifically "gldmHighGrayLevelEmphasis" and "glrlmHighGrayLevel RunEmphasis", as differentiators between coronary texture in patients with and without hypercholesterolaemia. These findings demonstrate the potential of advanced image analysis techniques in discovering relationships in the field of cardiovascular disease \cite{Kahmann2023}.
\subparagraph{}
With the gradual adoption of PCCT in clinical contexts and the increasing availability of research shedding light on this subject, there is growing interest in exploring the convergence of PCCT with the capabilities of deep learning algorithms in medical imaging. Therefore, this review aims to demonstrate how PCCT and deep learning can work together to improve medical imaging.

\subsection{Research Questions}
\paragraph{}
The main objective of this review is to evaluate recent articles on PCCT applications in deep learning and radiomics studies, published up to January 2024. Therefore, this review aims to answer the following questions: 1) Do the enhanced capabilities of the PCCT improve radiomics analysis? 2) How does PCCT differ from EICT in terms of radiomic analysis? 3) How does deep learning improve image quality in PCCT, specifically through image reconstruction and noise reduction techniques? 4) How is material decomposition performed in PCCT and is it still sensitive to image artifacts?

\subsection{Search Strategy} 
\paragraph{}
This review followed the PRISMA (Preferred Reporting Items for Systematic Reviews and Meta-Analyses) diagram \cite{Prisma}. 
The methodological approach involved a comprehensive analysis of PCCT technology, involving Deep Learning and Radiomics research papers. 
The search was performed in the following databases including PubMed, ScienceDirect and Scopus with two search queries '("Photon Counting CT" AND "deep learning")', and '("Photon Counting CT" AND "radiomics")' to find specific papers to access the several topics previously mentioned.
\subparagraph{}
The original search identified 1,949 articles, 64 in PubMed and 1,885 in Scopus. After removing duplicate (218), ineligible records (1,275) and not English (seven), 449 articles were identified for screening. Ineligible records included encyclopedias, book chapters, conference abstracts, discussions, and editorials. 
Screening of 449 articles based on the titles and abstract resulted in excluding another 386 articles. These articles were excluded because the abstracts did not contain information about PCCT or information related to it. In total 63 articles were intended for inclusion in this manuscript. Finally another 11 articles were not accessible and therefore this resulted in a total of 52 final papers regarding topics related to PCCT. The PRISMA \cite{Prisma} diagram in Figure 1 provides a summary overview of the screening.

\subsection{Manuscript outline}
\paragraph{}
This review explores how deep learning and radiomics are used in PCCT. It looks at recommendations from different studies and provides an overview of them, comparing PCCT with EICT to see which is considered better in each case. This section presents the outline of the manuscript. The rest of this review is organized as follows.
\begin{itemize}
\item
  \textbf{Section 2} discusses in detail the differences between PCCT and EICT in case studies applied to patients, and what each system allowed to conclude in each situation.
\item
  \textbf{Section 3} discusses the current techniques used in the radiomics study with PCCT, and how it improves the findings drawn from the radiomics study of PCCT medical images.
\item
  \textbf{Section 4} provides information on how various deep learning techniques in tasks such as segmentation, image reconstruction, material decomposition, artifact and noise reduction can further improve the properties of the medical image obtained with PCCT, presenting real cases and their conclusions.
\item
  \textbf{Section 5}  looks at what the future of PCCT will be like, given that it is a recent imaging modality.
\item
  Finally, \textbf{Section 6}, concludes on the topics explored in the previous sections, highlighting the research options that have already been pursued and obtained positive results.
\end{itemize}

\begin{figure}[ht]
    \centering
    \includegraphics[width=10cm, height=8cm]{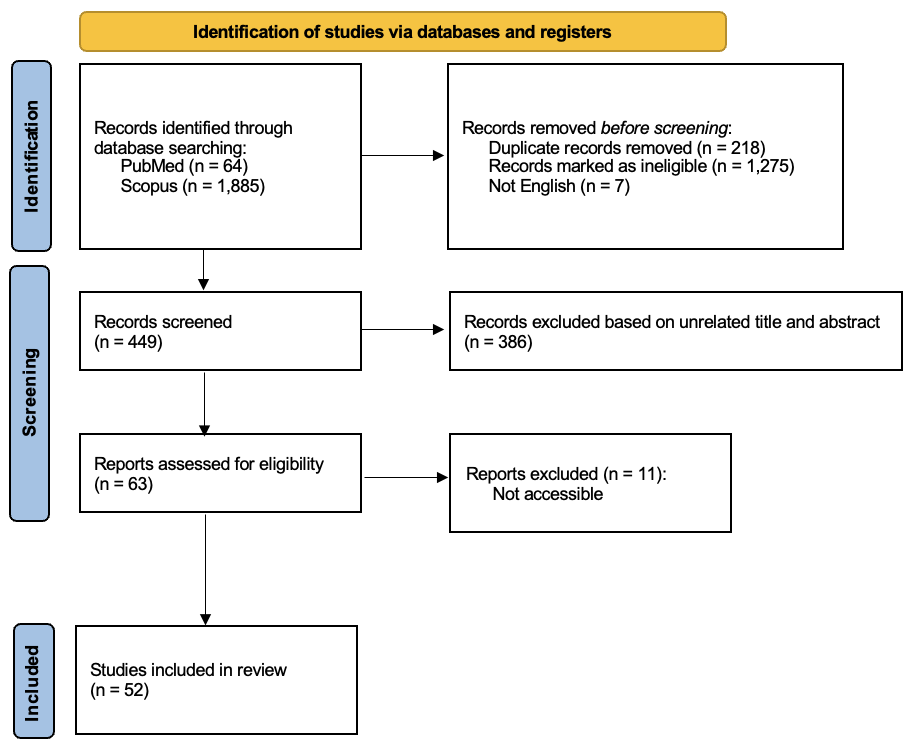}
    \caption{Selection of the papers according to PRISMA diagram \cite{Prisma}.}
    \label{fig:prisma review}
\end{figure}

\section{EICT and PCCT Systems}
\paragraph{}
PCT is a technique that will change the clinical use of CT in the future. Clinical CT scanners use EIDs, which differ from PCDs. To convert X-rays into electrical signals, most CT systems today use third-generation rotational designs and solid-state EIDs. The EID technology relies on the indirect conversion of X-ray photons into visible light, which is then captured by a layer of photodiodes. This light is then converted into an electrical output signal that correlates with the total energy deposited over a measurement interval. The electrical output signal of the detector does not provide information about the energy of individual photons because it integrates the energy from all photons. Figure 2 is a diagram of the PCD explained earlier. This way, PCDs can analyze and extract more information from the human body than the EICT.
\subparagraph{}
PCDs are capable of directly converting X-ray photons into electrical signals with a high-speed semiconductor. They can count the number of individual photons that exceed a predefined energy level. The design of a PCD involves a high atomic number semiconductor material, such as cadmium telluride (CdTe), gallium arsenide, or cadmium zinc telluride (CdZnTe), which covers the entire surface of the detector. When X-ray photons interact with the semiconductor, they generate electron-hole pairs, the number of pairs being proportional to the energy of the incident photon. The detector counts these pairs individually, allowing for precise photon counting. The electronic readout processes the signals, allowing spectral images and advanced artifact reduction techniques to be obtained \cite{Meloni2023_review}.
\subparagraph{}
When the X-ray photons interact with the semiconductor, this interaction creates a charge cloud of electron-hole pairs. 
An electric field is applied to the PCD by using an external electric voltage. This facilitates the movement of the charge cloud towards the pixel's electrodes, particularly the anode pixel, resulting in the generation of a pulse.
In an ideal scenario, a single photon produces a single charge cloud, which enters a pixel electrode perpendicularly and generates a single pulse. The size of the pixel electrode influences spatial resolution in PCCT scans, so reducing its size results in an improvement of spatial resolution. However, reducing the size of the pixel electrode increases the chance of measurement errors due to charge sharing and charge leakage.
The detector's electronics record the number of pulses whose amplitude exceeds a predefined threshold. This threshold is adjusted to be above the electronic noise level but below the pulses generated by the incident photons. In addition, by comparing each pulse with various threshold levels, the detector can categorize the incoming photons into different energy ranges (usually between two and eight), depending on their energy \cite{Meloni2023_review}, \cite{Nakamura2023}.
\subparagraph{}
Therefore, PCCT offers several advantages over traditional CT methods. These include a higher CNR, improved spatial resolution, and optimized spectral imaging. PCCT provides the ability to reconstruct images at a higher resolution, correct beam-hardening artifacts, and perform multi-energy/multi-parametric imaging based on the atomic properties of tissues. This allows for the use of different contrast agents and improves quantitative imaging. Furthermore, it facilitates radiation dose reduction through its ability to generate VNC images. These combined benefits contribute to enhanced image quality, reduced patient exposure to contrast agents and radiation, and improved safety in diagnostic procedures \cite{Review1_judith}, \cite{Review2_willemink}, \cite{Sartoretti2023_review}.

\begin{figure}[ht]
    \centering
    \includegraphics[width=1.0\linewidth]{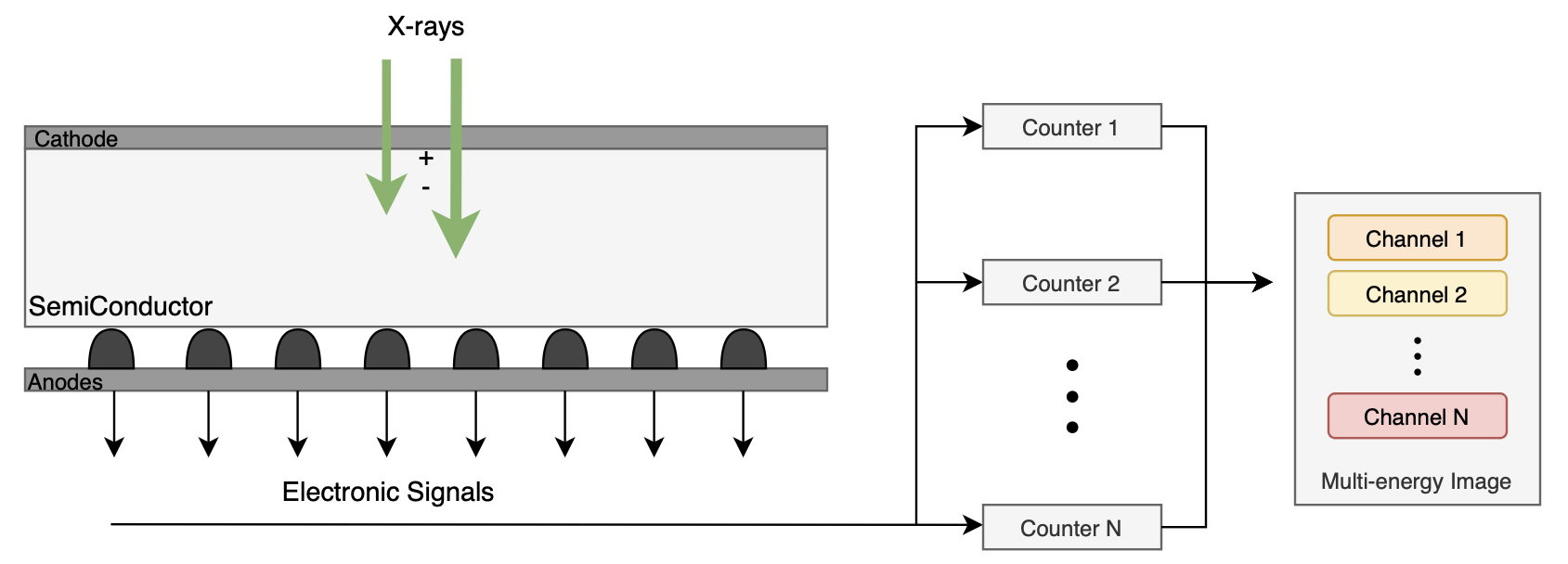}
    \caption{Schematic drawing of Photon Counting Detector. Adapted from \cite{Meloni2023_review}.}
    \label{fig:PCCT}
\end{figure}

\subparagraph{}
Table 1 shows a selection of scientific articles that prove the superiority of the image quality obtained through PCCT, highlighting its characteristics compared to conventional CT methods.
\subparagraph{}
Sharma et al. conducted a study using anthropomorphic computer models of various anatomical regions, including the lungs, liver, inner ear, and head and neck, with specific anomalies. These models were examined using a PCCT and a conventional EICT system with the corresponding dose levels. Radiologists then evaluated the resulting images to evaluate the perceptual advantages of PCCT compared to EICT in terms of anatomical features visualization and image quality. PCCT demonstrated superior image quality and improved visualization capabilities, resulting in better task-specific perceptual performance compared to EICT for most clinical tasks. Statistically significant improvements were observed in functions related to the visualization of lung lesions, liver lesions, soft tissue structures, and the cochlea of the inner ear. PCCT, therefore, showed improvements in task-specific perceptual performance compared to EICT, with statistically significant improvements in 6 out of 20 tasks. The research suggests that optimized spectral information and reconstruction kernels can significantly enhance the perceptual performance of PCCT. By improving image quality and visualization capabilities, these enhancements are expected to benefit routine clinical practice \cite{Sharma2023_silicon}.
\subparagraph{}
Rajagopal et al. conducted study comparing PCCT with EICT at various dose levels. Using an investigational scanner, image quality was compared at various dose levels, including 1.7, 2.0, 4.0, and 6.0 mGy CTDIvol, all below conventional abdominal CT doses. Phantoms were utilized for noise, resolution, CNR, and detectability index measurements. PCCT demonstrated a 22.1\%-24.0\% improvement in noise across dose levels, resulting in a 29\%-41\% improvement in CNR and a 20\%-36\% improvement in detectability index. Additionally, PCCT exhibited higher CNR for iodine contrast across all evaluated doses and iodine concentrations. PCCT displayed superior image quality compared to EICT, particularly advantageous for iodine detection in low-dose conditions \cite{Rajagopal2021}. Other investigations have performed similar research and reached consistent conclusions, confirming that the PCCT system allows for a substantial reduction in radiation exposure without compromising image quality and noise levels.
\subparagraph{}
Patzer et al. conducted a study to determine if PCCT, using the capabilities of UHR and without additional radiation, can provide a better representation of the trabecular microstructure and significantly reduce noise compared to EICT. In addition, the study aimed to evaluate PCCT as a promising alternative to EICT for the assessment of shoulder trauma in routine clinical practice. Sixteen cadaver shoulders were imaged using both scanners with 120 kVp acquisition protocols with corresponding doses. The PCCT scans, performed in UHR mode without the need for an additional post-patient comb filter, were contrasted with EICT scans following the clinical standard as "non-UHR". The reconstructions used different kernels. Radiologists subjectively assessed image quality, revealing that UHR-PCCT outperformed non-UHR-PCCT and EID-CT in subjective image quality. Remarkably, low-dose UHR-PCCT outperformed full-dose studies without UHR in any of the scanners. Quantitative analysis showed the lowest noise and highest SNR in non-UHR-PCCT reconstructions. This study concludes that PCCT in UHR mode provides superior imaging of trabecular microstructure without additional radiation dose. As shown in Figure 3, it is a promising alternative for the evaluation of shoulder trauma in clinical practice \cite{Patzer2023_nature}. Figure 3 was retrieved from \cite{Patzer2023_nature}.

\begin{figure}[ht]
    \centering
    \includegraphics[width=12cm, height=8cm]{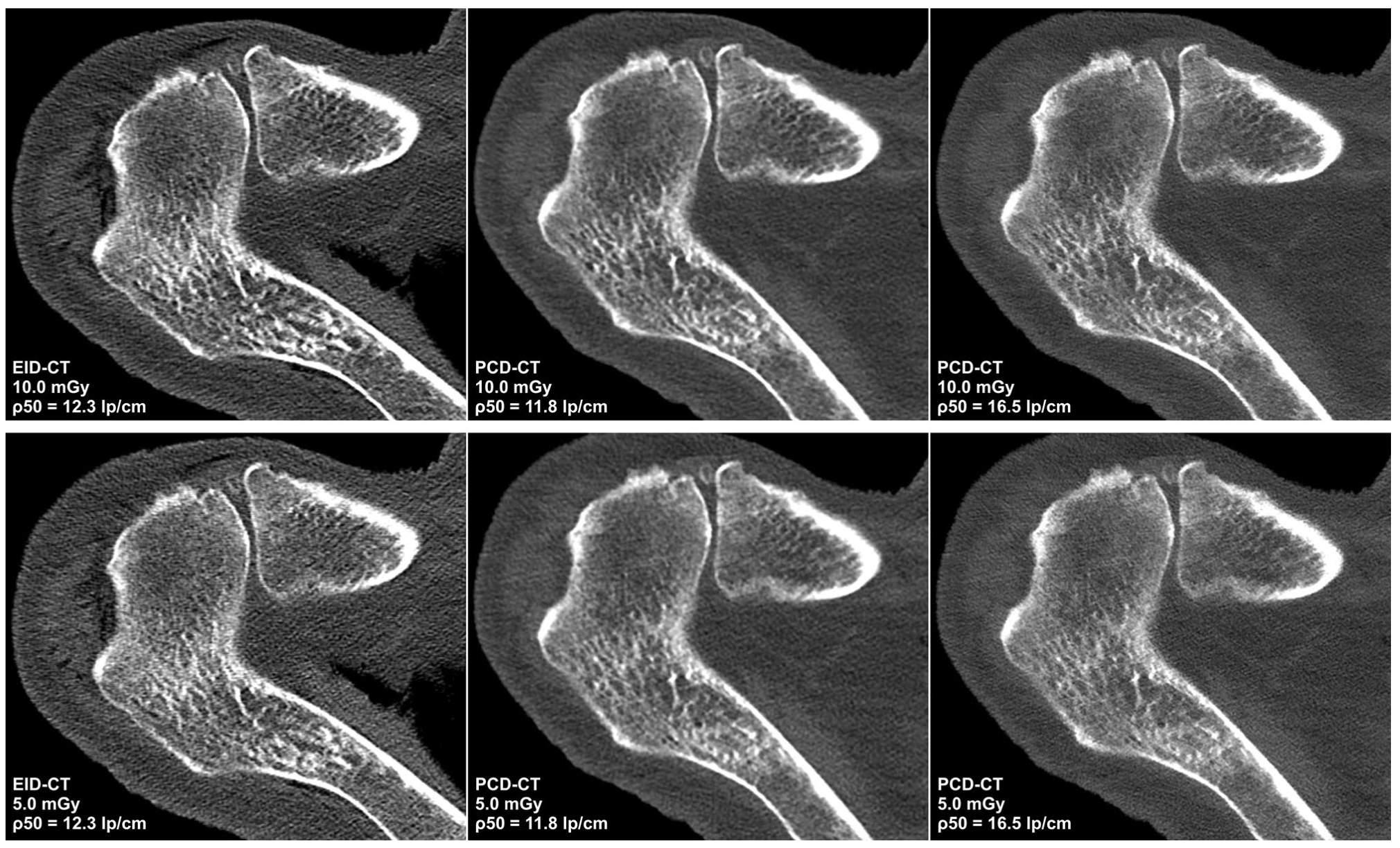}
    \caption{ Representative axial CT slices at the acromioclavicular joint level showcase image quality variations across six combinations of dose protocols and detector technologies. The upper row displays full-dose scan protocols (10.0 mGy) for EICT, non-UHR-PCCT, and UHR-PCCT, while the lower row exhibits low-dose scan protocols (5.0 mGy) for the same technologies. Notably, lower doses in EICT reveal increased image noise, potentially impacting the assessability of cancellous bone structures. Retrieved from \cite{Patzer2023_nature}.}
    \label{fig:shoulder}
\end{figure}

\subparagraph{}
Pourtmorteza et al. focused on the performance of PCCT in imaging the human brain, comparing its performance with the conventional EICT. Radiation dose-matched scans were conducted on an anthropomorphic head phantom and 21 human volunteers. PCCT demonstrated higher reader scores for gray-white matter differentiation and lower image noise, with excellent reproducibility. Quantitative analysis revealed 12.8\%-20.6\% less image noise, 19.0\%-20.0\% higher SNR, and 15.7\%-33.3\% higher contrast-to-noise ratios for PCCT. The findings suggest that photon-counting detector brain CT offers superior gray-white matter contrast, attributed to enhanced soft-tissue contrast and reduced image noise \cite{Pourmorteza}.

\begin{longtable}{|p{1.8cm}|p{2cm}|p{2.4cm}|p{2.7cm}|p{6cm}|}
\captionsetup{singlelinecheck=false, justification=raggedright}
\caption{Studies and respective clinical outcomes regarding the comparison between EICT and PCCT systems and image quality.}\\
    \hline
    \multicolumn{1}{|c|}{\textbf{Study}} &
    \multicolumn{1}{c|}{\textbf{Scans}} &
    \multicolumn{1}{c|}{\textbf{Purpose}} &
    \multicolumn{1}{c|}{\textbf{Metrics}} &
    \multicolumn{1}{c|}{\textbf{Metric Explanation}}\\
    \hline
    \endfirsthead
    \hline \multicolumn{5}{r}{{(Continues on next page)}} \\ 
    \endfoot
    \hline
    \endlastfoot
    Sharma et al. (2024) \cite{Sharma2023_silicon} & Anthropomor phic models for lungs, liver, inner ear, and head-and-neck & Assess the benefits of PCCT versus EICT for diverse clinical tasks and anatomical regions & Observer scores and Wilcoxon rank-sum test & - Mean observer scores evaluate the overall visualization performance and image quality;
    
    - The Wilcoxon rank-sum test identifiy statistically significant improvements for PCCT over EICT (lungs, liver, soft tissue structures, cochlea,and inner ear).\\
    \hline
    Rajagopal et al. (2021) \cite{Rajagopal2021} & Phantom & Explore the potential of PCCT to improve quantitative image quality compared to EICT & Noise and resolution measurements at various dose levels, CNR, detectability index and Iodine contrast measurements & - Noise measurements at different dose levels measure random variability or interference in the images;

    - Resolution assessments identify differences in the capability to provide clear and detailed images at different dose levels;

    - CNR compares the contrast and noise levels in the images obtained;

    - Detectability index evaluate how well PCCT and EICT detect specific structures or abnormalities under varied conditions;

    - Iodine contrast measurements examine the visibility and contrast of iodine-containing structures in the images to understand how well PCCT and EICT distinguish iodine contrast at various dose levels. \\
    \hline
    Pourmorteza et al. (2018) \cite{Pourmorteza} & Anthropomor phic head phantom and 21 human volunteers & Compare PCCT with EICT for human brains & Noise measurements, SNR, CNR, image quality score, paired t test, and the Wilcoxon rank-sum test & - Noise measures the amount of random variability present in images;
    
    - SNR assess how well distinguishable signals are from noise in images, providing insights into the image quality;
    
    - CNR measure the ability of the scans to differentiate between different tissues in the brain;
    
    - Image quality scores provide a subjective evaluation of the clarity and visual appeal of images;
    
    - Paired t Test and Wilcoxon Rank-Sum teste assess whether there are statistically significant differences between measurements obtained from the two types of CT scans. \\
    \hline
\end{longtable}
\newpage

\section{Radiomics Studies on PCCT images} 
\subsection{Radiomics Workflow}
\paragraph{}
Radiomics combines computer science, artificial intelligence, and radiology to improve the accuracy of medical imaging.
It is a method that extracts many features from medical images using data-characterization algorithms. In other words, radiomics quantifies texture information using artificial intelligence analysis methods by mathematically extracting the spatial distribution of signal intensities and pixel relationships \cite{Timmeren_Radiomics}.
\subparagraph{}
A typical radiomics study consists of numerous steps, each of which may be influenced by various factors. In the context of PCCT scans, the methodology typically involves the following key steps:
\begin{enumerate}
    \item \textbf{Imaging acquisition}
    \item \textbf{Image segmentation} - identify and delineate ROI in two dimensions or the VOI in three dimensions in PCCT scans. This often involves segmenting specific anatomical structures or lesions, which can be done manually with the help of radiologists, semi-automatically with standard image segmentation algorithms such as region growing or thresholding, or fully automatically with deep learning algorithms. Different open-source software solutions are available, such as 3D Slicer, ITK-SNAP, or ImageJ.
    \item \textbf{Image Processing} - attempts to homogenize the images from which the radiomics features will be extracted in terms of pixel spacing, grey-level intensities, and grey-level histogram bins. At this stage, it is essential to report each detail of the image processing step to ensure the reproducibility of the results.
    \item \textbf{Feature Extraction} - refers to the calculation of the characteristics used to quantify the intensity of gray levels (histogram), shape, texture, transformation and radial aspects within the ROI/VOI. It is recommended to follow the Imaging Biomarker Standardization Initiative guidelines due to the variety of methods used to calculate these features. PyRadiomics is a commonly used tool in this process. Radiomic features are of various types, such as intensity-based (histogram), shape-based, texture-based, transform-based and radial features. During feature extraction, different filters are often applied, such as wavelet or Gaussian filters.
    \item \textbf{Feature Selection} - identify the most informative features that significantly impact the task of diagnosis or prognosis through statistical analysis or cluster analysis.
\end{enumerate}

\subparagraph{}
In summary, the radiomics methodology for PCCT exams involves a systematic process of segmentation, image processing, feature extraction and feature selection. This final set of features is then used to create predictive models and extract valuable information to aid clinical decision-making, thus increasing the usefulness of PCCT scans in healthcare \cite{Timmeren_Radiomics}. 
Dunning et al. developed a ML model, a SVM to automatically classify the risk of coronary plaques based on radiomic features extracted from different types of images obtained by PCCT \cite{Dunning2023}, and Mundt et al. developed a logistic regression model that allows predicting the presence of calcification in coronary arteries by analyzing the features extracted by radiomics \cite{Mundt2023}. Landsmann et al. use a CNN-based approach to distinguish breast density, and therefore related health problems (higher tissue density, higher probability of breast cancer), through features derived from texture analysis \cite{Landsmann2022_breast}.
Table 2 summarizes the articles that study and analyze radiomics in PCCT.
\subparagraph{}
Mundt et al. investigated the relationship between the texture of periaortic adipose tissue, analyzed using radiomic features, and coronary artery calcification, since cardiovascular diseases are one of the main causes of global mortality, and therefore their understanding and identification remains crucial. 
The patients (mean age 56 years, 34 men, 21 women) underwent PCCT examinations. The periaortic adipose tissue surrounding the thoracic descending aorta was manually segmented on non-contrast axial images. Patients were stratified into three groups based on their coronary artery calcification status: Agatston score 0, Agatston score 1-99, and Agatston score $\geq$ 100. Pyradiomics was used to extract radiomic features (106 in total). R statistics facilitated the statistical analysis, including mean and standard deviation calculations, along with Pearson's correlation coefficient for feature correlation. Random Forest classification helped with feature selection, and visualization was achieved through boxplots and heatmaps.
The results highlighted two higher-order radiomic features, "glcm ClusterProminence" and "glcm ClusterTendency", which differed between patients without CAC and those with significant CAC according to Random Forest classification. "glcm ClusterProminence" emerged as the main differentiating feature. The results suggest that changes in the texture of periaortic adipose tissue may be correlated with coronary artery calcium score, suggesting a possible influence of inflammatory or fibrotic activity in perivascular adipose tissue \cite{Mundt2023}. 
\subparagraph{}
Other studies investigating radiomic analysis in PCCT have been performed, such as Tharmaseelan et al. who studied the metabolic activity of perivascular adipose tissue and its association with aortic calcification and aortic diameter. The study explores the potential influence of periaortic adipose tissue on the development of aortic calcification. Semi-automatic segmentation of the periaortic adipose tissue was performed and Pyradiomics was used to extract radiomic features. Statistical analysis in R statistics involved calculating the mean and standard deviation, along with Pearson's correlation coefficient for feature correlation. Random Forest classification was used for feature selection and an unpaired two-tailed t-test was applied to the final set of features. The results, presented as boxplots and heatmaps, covered data from 30 patients categorized based on the presence of local aortic calcification. Using Random Forest feature selection, a set of seven higher-order features was identified to differentiate periaortic adipose tissue texture between the two subgroups (with and without local aortic calcification). The subsequent t-test showed statistically significant discrimination for all characteristics. The results therefore suggest that textural changes in periaortic adipose tissue, associated with the presence of local aortic calcification, could potentially serve as the basis for identifying a biomarker for the development of atherosclerosis \cite{Tharmaseelan2022}.
\subparagraph{}
Staying focused on the heart region, Dunning et al. performed a study using PCCT images to classify coronary plaque risk using a ML model based on radiomics. Nineteen patients with coronary CT angiography were scanned on a PCCT system, generating five types of images: VMIs at 50, 70 and 100 keV, iodine maps and VNC. A total of 93 radiomic features were extracted from each image and a statistical analysis was carried out to identify significant features that distinguished low- and high-risk plaques. Two significant features were then input into a SVM for ML classification. The study used a LOOCV strategy, achieving classification accuracy. The results indicated that the ML model, particularly using 100 keV VMIs and VNC images, effectively differentiated between low- and high-risk coronary plaques in PCD-CTA \cite{Dunning2023}.

\begin{longtable}{|p{1.8cm}|p{1.8cm}|p{2cm}|p{3.4cm}|p{5.6cm}|}
\captionsetup{singlelinecheck=false, justification=raggedright}
\caption{Studies and respective clinical outcomes regarding the radiomics analysis on PCCT.}\\
    \hline
    \multicolumn{1}{|c|}{\textbf{Study}} &
    \multicolumn{1}{c|}{\textbf{Scans}} &
    \multicolumn{1}{c|}{\textbf{Purpose}} &
    \multicolumn{1}{c|}{\textbf{Metrics}} &
    \multicolumn{1}{c|}{\textbf{Metric Explanation}}\\
    \hline
    \endfirsthead
    \hline \multicolumn{5}{r}{{(Continues on next page)}} \\ 
    \endfoot
    \hline
    \endlastfoot
    Dunning et al. (2023) \cite{Dunning2023} & 19 patients & Investigate the differentiation between low and high-risk coronary plaques using the ML model developed & Manhattan plots and T-test, SRCC, and SVM & - Manhattan plots for each image type, followed by independent two-tailed t-tests on radiomic features comparing high- and low-risk plaques;
    
    - Characteristics that did not reach a significance level of p $<$ 0.05 were excluded;
    
    - SRCC is calculated between pairs of features, with selection based on the lowest p-values and SRCC $\leq 0.4 $. If SRCC $>$ 0.400 with other traits, the first one was discarded and the process was repeated;
    
    - Two chosen features were fed into an SVM trained with radiologists' reports as ground truth, using a LOOCV strategy. The accuracy of the SVM, measured by the number of correctly classified plaques in each LOOCV fold, assesses the risk stratification performance for each image type. \\
    \hline
    Mundt et al. (2023) \cite{Mundt2023} & 55 patients & Assess the correlation between periaortic adipose tissue texture and coronary artery calcification through radiomics analysis & Agatston Score, statistical analysis (Pearson's correlation coefficient), Random Forest (RF), Boxplots and heatmaps, Monovariable logistic regression (MLR) & - Agatston Score categorized patients into three groups (0, 1-99, $\geq$ 100) to explore the connection between periaortic adipose tissue texture and coronary artery calcification severity; 
    
    - Statistical analysis delved into feature relationships, while RF aided in selecting radiomics features distinguishing patients with and without calcification;
    
    - Boxplots visualized feature distributions, and heatmaps represented feature correlations; 
    
    - MLR predicted Agatston Score > 0 based on selected features, unraveling individual feature influences; 
    
    - Features underwent multicollinearity testing, and 10-fold cross-validation to assess the stability of the main feature. \\
    \hline
    Landsmann et al. (2022) \cite{Landsmann2022_breast} & 200 patients & Investigate the potential of texture analysis in distinguishing breast density in spiral breast PCCT & ANOVA test, Spearman's Rho, Cluster analysis, and MLR & - ANOVA compare the texture features for different levels of breast density with Bonferroni correction to retain the significant features;
    
    - Spearman's Rho evaluates the correlation between features and breast density;
    
    - Cluster analysis explores the dependencies between the features; 
    
    - MLR divides the data into training (70\%) and test (30\%) sets, calculating probabilities for each breast density category;
    
    - ICC measures the agreement in the assessment of breast density between readers.\\ 
    \hline
    Tharmaseelan et al. (2022) \cite{Tharmaseelan2022} & 30 patients & Study the relationship between the features of perivascular adipose tissue and aortic calcification for the development of atherosclerosis & Cluster analysis, Random Forest and two-tailed — unpaired t-test &  - Cluster analysis grouped patients based on radiomics features, visually comparing them in a Heatmap for abdominal aortic calcification presence;
    
    - Boruta/Random Forest identified seven crucial features distinguishing patients;
    - A two-tailed unpaired t-test confirmed significant differences for all features, visually represented through boxplots. \\
    \hline
\end{longtable}

\newpage
\subsection{Difference in Radiomic Analysis between PCCT and EICT}
\paragraph{}
The radiomics pipeline described above is unique for each situation and each image to be studied. Therefore, the main differences in radiomics analysis between PCCT and EICT result from the different imaging technologies, energy discrimination capabilities, and resolutions of these two CT approaches. PCCT's ability to provide more detailed, material-specific information can influence the radiomics features that can be extracted and, consequently, the depth of quantitative analysis possible. Table 3 presents several research articles showing the comparison between PCCT and EICT radiomic studies.
\subparagraph{}
The first studies, such as the studies by Dunning and Wolf, were performed on phantoms or organic samples. 
Dunning's study demonstrated that, in organic samples, 14 radiomics features were obtained in both scans. Still, when they were compared, it was found that 13 of the 14 radiomics features studied were visibly altered by the improved resolution of the PCCT system. These results show that the features derived from high-resolution PCCT can potentially alter radiomics-based clinical interpretation \cite{Dunning2022_impact}.
As a complement, Wolf et al. studied the impact of different VMI reconstructions on radiomic features in \emph{in vitro} and \emph{in vivo} PCCT datasets. 12 organ phantoms and 23 patients were scanned. In the phantom scans, images were reconstructed at six keV levels and the results showed excellent repeatability of the radiomic features. 
For \emph{in vivo} patients, the percentage of stable features was high among 90 keV \(\geq\) VMIs. At the same time, comparison with lower levels of VMI led to fewer reproducible features, so it was concluded that there is a better reproducibility correlation between higher VMIs in myocardial texture \cite{Wolf2023}.
\subparagraph{}
Sharma et al. conducted a study using an anthropomorphic phantom with three different morphologies for lung lesions, comparing the EICT with the radiomic study. Both systems extracted 14 radiomic features. The findings indicate a reduced mean estimation error in PCCT, specifically in independent features (35.9\% vs. 54.0\%) and overall features (54.5\% vs. 68.1\%). Statistically significant error reductions were observed for eight out of the fourteen features. In the experiments, tests were performed to see the effect of changing parameters such as the radiation dose, the distance to the isocenter, the pixel size and the reconstruction kernel. For both systems, the accuracy of the estimate was minimally affected by the radiation dose administered and the distance from the isocenter of the model used, while the reconstruction kernel and pixel size had a relatively stronger effect on the results obtained \cite{Sharma2023}.
\subparagraph{}
Another study compared the properties of the radiomic features of the myocardium on CT with an EID and a PCD. In a total of 50 patients, they obtained similar results with regard to first-order features. However, the higher-order features showed a partially significant difference between PCCT and EICT, being clearer with PCCT. This indicates a possible impact of better spatial resolution, better low-energy photon detection and a better SNR on texture analysis in PCCT \cite{Ayx2022}. On the same topic, the same author wrote a study with 30 patients in which it was possible to differentiate between patients with coronary artery calcification and patients without coronary artery calcification using four different myocardial texture parameters ("gldm SmallDependenceHighGrayLevelEmphasis", "glcm ClusterShade", "glrlm LongRunLowGrayLevelEmphasis", and "ngtdm Complexity") extracted using radiomics. Another proof of the importance of radiomics analysis in PCCT \cite{Ayx2022_myocardial}.
\subparagraph{}
Schwartz's study compared the image quality of high-resolution, low-dose PCCT with standard chest EICT in the same patients. After all radiomics pipeline procedures, results showed that PCCT produces higher measured and perceived image quality at lower radiation doses than EICT, despite greater measured quantitative and qualitative noise by using a lower dose of radiation \cite{SchwartzRia2023}.
\subparagraph{}
Regarding PCCT in oncology patients, Allphin applied transplanted soft tissue sarcomas with a high mutational load in Rag2+/- and Rag2-/- mice to model the variable load of lymphocytes. They were tested in EICT and PCCT, and the radiomics features were extracted. The radiomics features allowed differentiation between Rag2+/- and Rag2-/- tumors; in other words, tumors with different lymphocyte loads have slight modifications in their architecture that are not detected with conventional image-derived tumor metrics but which are better revealed in radiomics analysis of PCCT images with nanoparticle contrast \cite{Allphin2022}.
\subparagraph{}
In differentiation tasks, Kahmann studied the differences in the presence or absence of hypercholesterolemia in the radiomics analysis of 66 patients and verified the existence of two radiomics features. Patients with hypercholesterolemia had a higher concentration of high-density values. Analysis of pericoronary adipose tissue texture discriminated between patients with and without hypercholesterolemia. This is an important step in the development of new treatments for cardiovascular diseases \cite{Kahmann2023}.

\begin{longtable}{|p{2cm}|p{1.4cm}|p{2.4cm}|p{3.4cm}|p{5cm}|}
\captionsetup{singlelinecheck=false, justification=raggedright}
\caption{Studies and respective clinical outcomes regarding the differences in radiomics analysis between PCCT and EICT.}\\
    \hline
    \multicolumn{1}{|c|}{\textbf{Study}} &
    \multicolumn{1}{c|}{\textbf{Scans}} &
    \multicolumn{1}{c|}{\textbf{Purpose}} &
    \multicolumn{1}{c|}{\textbf{Metrics}} &
    \multicolumn{1}{c|}{\textbf{Metric Explanation}}\\
    \hline
    \endfirsthead
    \hline \multicolumn{5}{r}{{(Continues on next page)}} \\ 
    \endfoot
    \hline
    \endlastfoot
    Dunning et al. (2022) \cite{Dunning2022_impact} & 16 organic phantoms & Evaluate the impact of the improved spatial resolution of PCCT, compared to EICT systems, on the calculation of radiomics features & The number of radiomics features, the noise quantification, the different dose levels, the reconstruction kernels, and the cluster analysis( Dunn index and Davies-Bouldin index) & - The number of features evaluates the impact of improved spatial resolution (both CT systems calculated 14);
    
    - Noise measures the standard deviation of CT numbers in an ROI (at different doses), different dose levels and reconstruction kernels explore the stability of radiomic features;
    
    - Dunn's index and Davies-Bouldin index quantify the separation of groups. \\
    \hline
    Wolf et al. (2023) \cite{Wolf2023} & 12 organic phantoms and 23 patients & Assess the impact of different VMI configurations on radiomics features in \emph{in vitro} and \emph{in vivo} PCCT datasets & ICC and Post Hoc Paired Samples ANOVA t-Test, percentage of stable features, VMI levels, statistical analysis (intra and inter-group comparisons) & - ICC assesses the repeatability and reproducibility of radiomic features;
    
    - Post Hoc paired samples ANOVA t-test compares radiomic features between various VMI reconstructions;
    
    - Percentage quantifies the proportion of radiomic features that remain stable in different VMIs, the different VMI levels assess their influence on radiomic features;
    
    - Statistical analysis analyzes the differences in radiomic features. \\
    \hline
    Sharma et al. (2023) \cite{Sharma2023} & Anthropo morphic phantom with three lung lesions morphologies & Compare the impact of silicon-based PCCT with conventional EICT on the estimation of morphological radiomics features in lung lesions & Estimation Errors, statistical significance and accuracy & - Estimation errors verify the accuracy of the estimated morphological radiological characteristics compared to the real values; 
    
    - Statistical significance (Wilcoxon test) validates the improvements observed in the PCCT;
    
    - Accuracy studies under various conditions (dose, reconstruction method, pixel size and distance from the isocenter) to understand the factors influencing both systems. \\
    \hline
    Ayx et al. (2022) \cite{Ayx2022} & 50 patients & Compare the radiomics properties of the left ventricular myocardium between PCCT and EICT & T-Test and F-Test, Fisher's exact test, Pearson Correlation Coefficient, heatmaps and statistical thresholds & - T-test for unpaired samples compares means of quantitative variables, while the F-test evaluates significant differences in variance. Both tests help researchers determine whether the differences were statistically significant;
    
    - Fisher's exact test analyzes associations between categorical variables, providing information on the significance of these associations; 
    
    - Pearson's correlation coefficient calculates the correlations of characteristics between quantitative variables (value 1 or -1: strong correlation; value close to 0: no linear relationship). \\
    \hline
    Ayx et al. (2022) \cite{Ayx2022_myocardial} & 30 patients & Identify and analyze radiomics features that may change in response to the severity of coronary artery calcification & Mean and standard deviation, Pearson correlation coefficient, boxplots and heatmaps, and random forest &  - Mean and standard deviation makes a quick analysis of the data;
    
    - Pearson's correlation coefficient helps to understand the correlations of features calculated and visualized in boxplots and heatmaps;
    
    - The RF feature selection defines a number of features (four) to discriminate between the two groups with regard to myocardial texture;
    
    - The Agatston score categorizes patients into subgroups based on the severity of coronary artery calcification. \\
    \hline
    Schwartz et al. (2023) \cite{SchwartzRia2023} & 60 patients & Compare the image quality, noise levels, and overall diagnostic performance of PCCT with standard EICT in the same group of patients & Subjective metrics evaluated by blinded radiologists (subsegmental bronchial wall definition, noise, overall image quality); Quantitative metrics: GNI, NPS parameters (fav, fpeak), TTF parameters (f10, d'adj) and applied output radiation doses (CTDIvol) & - Subjective metric rates on a scale of 0 (worst) to 100 (best) to assess the clarity and definition of the subsegmental bronchial walls, the perceived level of noise or granularity and the overall diagnostic quality of the images (resolution, contrast and artifacts); 
    
    - Quantitative metrics: GNI assesses the average noise level, fav analyzes the average frequency content of the noise, fpeak identifies the frequency at which noise in CT images is most pronounced, f10 indicates the frequency at which 10\% of the task (e.g. bronchial wall details) is transferred, d'adj measures the detectability of features and CTDIvol assesses the amount of radiation delivered to the patient during the CT examination. \\
    \hline
    Allphin et al. (2022) \cite{Allphin2022} & 25 rats & Investigate whether radiomics analysis, based on micro PCCT with nanoparticle contrast enhancement differentiate tumors based on lymphocyte burden & Univariate and multivariate analysis and to evaluate performance AUC, accuracy, precision and recall & - Univariate analysis evaluates the association between each radiomic feature and the genotype;
    
    - Multivariate analysis uses the MRMR algorithm for feature selection;
    
    - Logistic Regression Binary Classifier trains subsets of features (EID, PCD, material maps) and Monte Carlo repeated stratified cross-validation evaluates the performance of the classifier;
    
    - AUC, accuracy, precision and recovery indicate the classifier's ability to discriminate, measure the overall correct predictions made by the classifier and demonstrate the classifier's ability to identify positive cases. \\
    \hline
    Kahmann et al. (2023) \cite{Kahmann2023} & 66 patients & Investigate how hypercholesterolemia influences pericoronary adipose tissue texture & Pearson correlation coefﬁcient, boxplots and heatmaps, Random forest & - Pearson's correlation coefficient indicates the intensity and direction (positive or negative) with which radiomic features are correlated with the presence or absence of hypercholesterolemia;
    
    - Boxplots show the dispersion and central tendency of radiomic features for different groups, and heat maps provide a visual representation of the correlation patterns between features;
    
    - RF selects the most influential features in distinguishing between the two groups. \\
    \hline
    Allphin et al. (2023) \cite{Allphin2023_mouse} & 123 rats & Evaluate cardiac function in mouse models expressing different APOE genotypes using PCCT & ICC, Post hoc Paired Samples ANOVA t Test, CNN segmentation metrics (Dice Score, Precision, Recall, and AUC), Pearson Correlation Coefficient, ANOVA, Cohen F Effect Sizes, Confidence Intervals, and Adjusted Post hoc results & - ICC assess the stability of the radiomic features;
    
    - Post hoc paired samples ANOVA t-tests determine significant differences in radiomic features; 
    
    - CNN segmentation metrics, including Dice scores, precision, recall and AUC, assess left ventricle segmentation performance;
    
    - Pearson Correlation Coefficient help researchers on the relationships between cardiac measurements;
    
    - Linear models identify differences between groups, assessing the effects and interactions of various factors on cardiac measurements.\\ 
    \hline
\end{longtable}

\subsection{Radiomics Limitations}
\paragraph{}
Although radiomics has shown its potential in various investigations for diagnosis, prognosis, and prediction, this field faces several challenges. The gap between available knowledge and clinical requirements often results in studies with limited practical applicability. When addressing clinically relevant questions, the reproducibility of radiomics studies is often compromised due to the lack of standardization, adequate reporting, or restrictions related to code and open-source data \cite{Timmeren_Radiomics}.
\subparagraph{}
Therefore, one of the main limitations of radiomics analysis is the reproducibility of radiomics features. To overcome this issue, a radiomics analysis was performed on phantom scans using a PCCT. The study aimed to assess the stability of radiomics analysis, focusing mainly on the consistency of the extracted radiomics parameters. The research revealed that most of the extracted radiomics features showed excellent stability, both in test-retest analyzes and in new analyzes after repositioning \cite{Hertel2023}. 
A similar study was carried out on human subjects and came to the same conclusion. These results highlight the robustness of radiomics analysis using PCCT data, marking a significant step towards integrating radiomics analysis into clinical practice \cite{Tharmaseelan2022_stability}. These articles are summarized in Table 4.

\begin{longtable}{|p{2cm}|p{1.7cm}|p{2.4cm}|p{3cm}|p{5.3cm}|}
\captionsetup{singlelinecheck=false, justification=raggedright}
\caption{Studies and respective clinical outcomes regarding the stability of radiomics on PCCT.}\\
    \hline
    \multicolumn{1}{|c|}{\textbf{Study}} &
    \multicolumn{1}{c|}{\textbf{Scans}} &
    \multicolumn{1}{c|}{\textbf{Purpose}} &
    \multicolumn{1}{c|}{\textbf{Metrics}} &
    \multicolumn{1}{c|}{\textbf{Metric Explanation}}\\
    \hline
    \endfirsthead
    \hline \multicolumn{5}{r}{{(Continues on next page)}} \\ 
    \endfoot
    \hline
    \endlastfoot
    Hertel et al. (2023) \cite{Hertel2023} & 16 organic phantoms & Evaluate the stability of radiomics analysis on PCCT & CCC, ICC, and Random forest analysis & - CCC evaluates the agreement between two sets of measures;
    
    - ICC measures the reliability of the measurement. It helped evaluate stability through test and retest and with various mAs values between phantom groups;
    
    - Random Forest analysis determined the importance of radiomics features in the distinction between phantom groups.  \\
    \hline
    Tharmaseelan et al. (2022) \cite{Tharmaseelan2022_stability} & 10 patients & Investigate the potential of spectral reconstructions from PCCT in addressing challenges related to feature stability and standardization in radiomics & CCC, ICC and ANOVA test & - CCC assesses reproducibility between different segmentations;
    
    - ICC evaluates consistency and stability of radiomics features; 
    
    - ANOVA compares average ICC values between different ROI sizes and dimensionality, helping to identify the most stable segmentation approach. \\
    \hline
\end{longtable}

\section{Deep learning applications in PCCT}
\paragraph{}
PCCT introduced high-resolution imaging but faces challenges such as quantum noise and metal artifacts. Deep learning comes into play - a transformative approach that takes advantage of AI. Deep neural networks are excellent at reconstructing raw photon count data into clearer, more accurate images. In addition to noise reduction and artifact correction, deep learning improves PCCT through super-resolution techniques, promising unprecedented spatial detail. This synergy addresses technical challenges and points to a future in which PCCT offers greater diagnostic accuracy and less radiation exposure, reshaping the landscape of medical imaging. Below are some of these deep learning application topics.

\subsection{Image Reconstruction Techniques}
\paragraph{}
CT image reconstruction is a mathematical procedure that produces tomographic images using X-ray projection data collected from various angles around the patient. This process significantly influences the image quality and, consequently, the radiation dose. Ideally, with a specified radiation dose, the aim is to reconstruct images with minimal noise while maintaining accuracy and spatial resolution. Improving image quality through reconstructions can reduce radiation dose since it is possible to reconstruct images of comparable quality with lower doses \cite{Graafen2023}.
\subparagraph{}
The first CT scanners in the 1970s used iterative reconstruction algorithms, but their clinical use was limited due to computational constraints. It wasn't until 2009 that the first commercial iterative reconstruction algorithms replaced conventional Filtered Back Projection (FBP). Reconstruction algorithms for CT vary in complexity, including hybrid, model-based, and fully iterative methods:
\begin{enumerate}
    \item \textbf{Iterative Reconstruction:} combines forward and backward projection steps to iteratively refine the image, for example, IRIS (Iterative Reconstruction in  Image Space);
    \item \textbf{Hybrid Approach:} combines the speed of FBP with the refinement capabilities of iterative reconstruction. The initial steps often involve FBP, followed by iterative refinements to improve the image to take advantage of both methods. An example is the iDose4 algorithm;
    \item \textbf{Model-based:} uses a statistical model to represent the image and the data acquisition process. It improves image quality through iterative refinement and complex optimization algorithms, providing greater spatial resolution and lower noise levels, like SAFIRE (Sinogram Affirmed Iterative Reconstruction).
\end{enumerate}

\subparagraph{}
Due to the additional multi-energy information and detector elements, PCCT data is more complex than conventional CT data. Consequently, specialized reconstruction algorithms are required. Many authors choose DL methods based on CNN to learn the mapping between images and data. This removes the need for explicit knowledge of the underlying statistical model. X-ray exposure per projection or the number of projections is minimized in order to reduce the image dose. However, this reduction leads to a lack of photons or undersampling artifacts in the image. In other words, the higher the radiation dose (the more harmful to the patient), less noise in the medical image \cite{Nadkarni2023}.
\subparagraph{}
Previous research has shown that noise reduction based on deep learning can provide performance equivalent to that of iterative reconstruction and a significant reduction of computing time. 
Several DL architectures have been used to remove noise in PCCT images, including U-networks \cite{Nadkarni2023}, \cite{Wang2023_noise2noise}, \cite{Niu2023}, Pie-Net \cite{Chang2023_pie}, encoder-decoder CNN \cite{Gong2021_CNN}, many models using CNN but then it varies inside it like the one that uses
image-based noise insertion and PCCT images to train a U-Net via supervised learning \cite{Huber2022}, or just a simple CNN \cite{Baffour2023}, or even a more extensive system involving CNNs named PKAID-net \cite{Chang2023_PKAID}. 
\subparagraph{}
Another approach to image reconstruction in PCCT is QIR algorithms, which have been developed to satisfy the increasing hardware and software requirements of PCCT in terms of data complexity, spectral information, and noise profile. There are four strength levels of QIR (1-4) \cite{Huflage2023_QIR1}.
Recent research, including studies by Woeltjen and Sartoretti, has shown the impact of the QIR algorithm on enhancing the overall quality of lung images in PCCT. This reinforces the algorithm's significance in medical imaging \cite{Sartoretti2022_QIR2}, \cite{Woeltjen2022_QIR4}, \cite{Gaillandre2023_QIR3}.
\subparagraph{}
Yang et al. introduced an advanced CT spectral reconstruction method focusing on reducing radiation doses and resolving data noise. The approach uses the inherent correlation between channels acquired from the same object to improve reconstruction quality. By employing two iterative reconstructions with sparsity and low-degree dual constraints, the authors designed a heuristic strategy to address the optimization challenges. They incorporated a fusion framework using guided filters to fuse the results, obtaining satisfactory visualization, higher PSNR and SSIM values, and reduced material decomposition errors \cite{Yang2023}.
\subparagraph{}
Gruschwitz et al. evaluated the influence of various vascular reconstruction kernels on the image quality of CT angiographies. The investigation incorporated a 1st generation PCCT and was matched with dose equivalent scans performed on a 3rd generation EICT. In particular, the reconstructions were meticulously carried out with various vascular cores, with a specific emphasis on aligning the individual modulation transfer functions between the two scanners. To assess the objective quality of the image, SNR and CNR were calculated. In addition, subjective assessments were made by six radiologists using a forced-choice pairwise comparison tool. An imperative characteristic to be analysed was intraluminal attenuation.
The results of the study emphasised that the use of sharper convolution kernels in PCCT consistently resulted in superior image quality when compared to EICT, regardless of the specific reconstruction kernels used. This implies that PCCT has advantages in terms of intraluminal attenuation, noise reduction and overall image quality in the context of angiographies. The better subjective image quality obtained with PCCT further facilitates the application of sharper convolution kernels, thus potentially improving the assessment of vascular structures in clinical contexts \cite{Gruschwitz2023}.

\subparagraph{}
Graafen et al. explored the application of PCCT for the non-invasive diagnosis of hepatocellular carcinoma (HCC). The research focused on assessing the improvements in image quality, including noise reduction and higher spatial resolution, offered by PCCT. The investigation involved phantom experiments to analyze different reconstruction kernels and a patient population study for HCC imaging. Results showed that softer reconstruction kernels (in image processing, a kernel refers to a small matrix used for various operations, such as blurring, sharpening, relief, and edge detection) yielded superior overall quality for HCC evaluation in PCCT, exhibiting the highest contrast-to-noise ratio in all contrast phases. There were no significant differences in image contrast or the lesion between the body and quantitative nuclei with equal sharpness levels.
The findings support the preference for softer reconstruction kernels in HCC assessment with PCCT, with quantitative kernels showing potential for spectral post-processing without compromising image quality \cite{Graafen2023}.
\subparagraph{}
However, despite advances in DL-based denoising, significant challenges remain. One notable concern is the potential for overfitting to training data, resulting in limited generalization performance. The achievement of robust generalization is particularly critical in removing noise from PCCT images. This is due to the variations in PCD energy thresholds influenced by different contrast agents, thus affecting noise levels in various energy channels. Moreover, there is a clear need for reliable evaluation metrics to assess the performance of DL-based noise reduction algorithms accurately. Table 5 below shows some of the articles mentioned above.

\begin{longtable}{|p{2cm}|p{1,6cm}|p{2.2cm}|p{3.2cm}|p{5.3cm}|}
\captionsetup{singlelinecheck=false, justification=raggedright}
\caption{Studies and their clinical results on image reconstruction approaches in PCCT.}\\
    \hline
    \multicolumn{1}{|c|}{\textbf{Study}} &
    \multicolumn{1}{c|}{\textbf{Scans}} &
    \multicolumn{1}{c|}{\textbf{Purpose}} &
    \multicolumn{1}{c|}{\textbf{Metrics}} &
    \multicolumn{1}{c|}{\textbf{Metric Explanation}}\\
    \hline
    \endfirsthead
    \hline \multicolumn{5}{r}{{(Continues on next page)}} \\ 
    \endfoot
    \hline
    \endlastfoot
    Grafen et al. (2023)\cite{Graafen2023} & Phantom and 24 patients & Evaluate and optimize imaging parameters for HCC imaging using three-phase PCCT of the liver & CNR, edge sharpness, qualitative analyses by raters, comparison of kernels (image quality criteria, noise, and overall assessment)  & - CNR measures the relationship between the contrast of an object and the background noise in the image;

    - Edge sharpness assesses how well-defined and clear the boundaries of structures are in the image;

    - Qualitative analyses involve subjective evaluations by expert raters, typically assessing visual aspects of the images.

    - Image quality criteria evaluates the overall quality of images based on predefined criteria (contrast, sharpness, and clarity).\\
    \hline
    Nadkarni et al. (2023) \cite{Nadkarni2023} & Vivo and ex-vivo mice and phantoms & Exploring the challenges of PCCT noise by introducing UnetU, a fast and efficient DL model & RMSE for test set reconstructions, SSIM and PSNR at different sub-sampling factors, comparing their performance with non-local multi-energy averaging (NLM) and UnetwFBP, both quantitatively and qualitatively & - RMSE evaluates how well UnetU approximates the iterative reconstruction by quantifying the differences between the model's predictions and the actual test set reconstructions;

    - Evaluating SSIM and PSNR at different sub-sampling factors helps assess how well UnetU maintains image quality and structural information as the data is undersampled. Higher SSIM and PSNR values suggest better preservation of image details;

    - Comparing performance with NLM and UnetwFBP involves assessing how UnetU compares with alternative methods, both quantitatively and qualitatively.\\
    \hline
    Sartoretti et al. (2022) \cite{Sartoretti2022_QIR2} & 52 patients & Determine the optimal strength levels of QIR for PCCT of the lung & NPS and TTF, GNI, and quantitative image quality metrics (noise levels on different levels of QIR), Global Signal-to-Noise Ratio Index (GSNRI), mean attenuation of lung parenchyma and subjective frading by readers & - NPS and TTF assess spatial resolution and noise amplitude characteristics, providing insights into how the algorithm affects the frequency content and spatial resolution of the reconstructed images;

    - GNI analyze noise levels in HU across different strength levels of QIR;
    
    - GSNRI assesses the balance between the signal (useful information) and noise in the images across different QIR levels;
    
    - Mean attenuation of lung parenchyma focuses on the average attenuation values of lung parenchyma in HU;

    - Human readers subjectively assess the images in terms of overall quality, image sharpness, and perceived image noise.\\
    \hline
    Woeltjen et al. (2022) \cite{Woeltjen2022_QIR4} & 29 patients & Examines the impact of a PCCT QIR on image quality & Subjective assessment by radiologists, image sharpness, SNR, effective radiation dose, and statistical significance tests & - A qualitative evaluation using a 5-point Likert scale carried out by three senior radiologists to assess the overall perceived quality of the images;

    - SNR provides an objective assessment of the clarity and intensity of the signal in relation to the noise;

    - Calculation of the effective radiation dose for PCCT and EICT examinations, allowing a quantitative comparison of radiation exposure between the two modalities.

    - Determination of statistical significance, including p-values, to assess differences in image quality metrics between PCCT and EICT scans.\\
\end{longtable}

\subsection{Segmentation Applications}
\paragraph{}
Segmentation algorithms play a crucial role in analyzing and interpreting PCCT scans. These algorithms aim to precisely identify and delineate specific organs, tissues, or lesions within the scanned images. By pixel-wise classification, segmentation algorithms capture the shapes and volumes of target structures, providing valuable information for diagnosis, treatment planning, and interventions. 
\subparagraph{}
These algorithms have been widely developed and improved with the advancements in DL methods. By automating the segmentation process, these algorithms eliminate the need for manual intervention, saving time and effort for radiologists and improving workflow efficiency. Segmentation algorithms in PCCT scans have been implemented using various approaches, such as automated region-growing algorithms \cite{Klintström_segmentation1} and CNNs \cite{Allphin2023_mouse}; besides, manual segmentation remains a viable option.

\subsection{Reduction of artifacts and noise}
\paragraph{}
Deep learning algorithms demonstrate a remarkable ability to reduce noise in medical images, improving overall image quality and boosting diagnostic confidence. Specifically, in scenarios such as low-dose CT scans characterized by inherently high noise levels, these models can be trained to effectively attenuate noise while maintaining crucial image details. The ability to reduce noise is especially advantageous for patient groups that have a higher sensitivity to radiation exposure, including pediatric patients or individuals undergoing frequent imaging.
In addition, spectral information from PCCT scans can be use to create innovative metal artifact reduction techniques. These techniques can integrate contrast from low-energy compartments and metal artifact reduction from high-energy compartments, resulting in CT images with significantly reduced metal artifacts \cite{Mese2023_synergy}. 
\subparagraph{}
Do et al. evaluated the efficiency of energy thresholds and acquisition modes in minimizing metallic artifacts in PCCT and EICT. The study found that PCCT has higher noise levels than EICT regardless of the tube's potential. This indicates that using high-energy compartments implies a trade-off between increased noise and decreased overall image contrast since the reconstruction process does not incorporate low-energy photons \cite{Do2020_artifacts1}. 
\subparagraph{}
The challenge of noise reduction is inherent in most deep learning tasks, as networks must adapt and generalize their performance to training data that contains noise. In the context of noise reduction in PCCT, there is a correlation between the radiation dose - potentially harmful to patients - and the level of noise in the reconstructed images. To reduce radiation exposure, it is necessary to reduce either the X-ray exposure per projection or the number of projections. However, this reduction leads to challenges such as photon starvation or image subsampling artifacts \cite{Gong2020}. 
\subparagraph{}
Zhou et al. demonstrated the utility of high-energy compartments in reducing metal artifacts induced by implants in various body parts, such as the spine, shoulder, wrist, ankle, and elbows. Their study compared EICT with PCCT with tin filtering, observing reduced metallic artifacts, greater diagnostic confidence, and better anatomical recognition in images within the > 75 keV bin, emphasizing the importance of the PCCT use in medicine \cite{Zhou2019_artifacts}.
Another study on the same topic was carried out by Anhaus et al. using multiple phantoms, comparing monoenergetic images from 40 keV to 190 keV (in 10 keV increments) with and without iMAR. Their conclusions indicated that mono-energetic imaging alone did not effectively reduce metallic artifacts in materials with high atomic numbers, such as dental implants, the head of the hip, and the embolization coil. However, the inclusion of iMAR demonstrated a significant reduction in metal artifacts. Specifically, the ideal keV was identified for spinal implants as 100 keV and 120 keV when scanning with 120 and 140 kVp with tin filtering, respectively \cite{Anhaus2022}.
\subparagraph{}
PCCT allows the reconstruction of mono-energetic images. Layer et al. studied polychromatic and monochromatic images with energy levels of 100 keV, 130 keV, 160 keV, and 190 keV, with and without iMAR, for unilateral and bilateral total hip arthroplasty patients. The researchers observed a reduction in metal artifacts and better diagnostic interpretation for bone, muscle, vessels, bladder, and soft tissue in polychromatic images with iMAR. They also found similar benefits in mono-energetic images ranging from 100 to 190 keV, with and without iMAR, with the best result observed in 100 keV mono-energetic images with iMAR \cite{Layer2023}.
Björkman et al. conducted an exhaustive evaluation, both subjective and objective, comparing the efficiency of PCCT and DECT with and without iMAR in the attenuation of metallic artifacts in a bovine knee with orthopedic plates and screws. The subjective evaluation by the researchers revealed higher scores for image quality and the assessment of fracture lines, bone structures, and the articular surface in images reconstructed with PCCT when compared to those obtained from DECT. Nevertheless, the objective analysis did not identify any statistically significant differences between the two imaging techniques \cite{Björkman}.
\subparagraph{}
In another area of application, Patzer et al. observed a reduction in metallic artifacts caused by dental implants and better visualization of soft tissues. They compared poly-energetic and mono-energetic images from 40 to 190 keV with iMAR and the same range without iMAR. In addition, they observed a more significant reduction in metallic artefacts and better visualization of soft tissue in mono-energetic images from 110 to 190 keV with iMAR, compared to poly-energetic images with iMAR \cite{Patzer2023}.
\subparagraph{}
These results contribute to understanding the optimization of imaging parameters to reduce metallic artifacts in CT scans, particularly with emerging technologies such as PCCT. Efficient denoising not only improves the detectability of smaller lesions and anatomical structures, but also plays a vital role in maintaining the reliability of quantitative imaging metrics. These metrics are crucial for monitoring therapy progress and evaluating treatment responses. Table 6 below shows some of the articles mentioned above regarding the material decomposition application of PCCT.

\begin{longtable}{|p{2cm}|p{1,6cm}|p{2.2cm}|p{3.2cm}|p{5.3cm}|}
\captionsetup{singlelinecheck=false, justification=raggedright}
\caption{Studies and respective clinical outcomes regarding the material decomposition application of PCCT.}\\
    \hline
    \multicolumn{1}{|c|}{\textbf{Study}} &
    \multicolumn{1}{c|}{\textbf{Scans}} &
    \multicolumn{1}{c|}{\textbf{Purpose}} &
    \multicolumn{1}{c|}{\textbf{Metrics}} &
    \multicolumn{1}{c|}{\textbf{Metric Explanation}}\\
    \hline
    \endfirsthead
    \hline \multicolumn{5}{r}{{(Continues on next page)}} \\ 
    \endfoot
    \hline
    \endlastfoot
    Do et al. (2020) \cite{Do2020_artifacts1} & Phantom and Cadaver & Identify optimal parameters for achieving the best image quality in the presence of metal implants, such as a hip prosthesis & Artifact percentage and volume, distance-dependent trend, differences in CT numbers and CNR and noise levels & - Artifact percentage assess metal artifacts presence in reconstructed images, calculating the ratio of the artifact-affected region to the total VOI;
    
    - Artifact volume quantifies the volume occupied by metal artifacts in the reconstructed images;

    - Distance-Dependent Trend captures the relationship between artifact percentage and the distance of the evaluated VOI from the prosthesis;

    - Differences in CT Numbers compares CT numbers between different acquisition modes to understand variations in attenuation values under various settings;

    - CNR evaluates the image quality by assessing the contrast between different tissues relative to the image noise;

    - Noise levels provides insights into image quality, higher noise levels may impact diagnostic accuracy. \\
    \hline
    Gong et al. (2020) \cite{Gong2020} & Phantom and animal datasets & Develop a CNN (Incept-net) to directly estimate material density distribution from multi-energy CT images, without relying on conventional material decomposition methods & MAE, noise metrics, total mass density and efficiency of the model performance & - MAE quantifies the accuracy of the predicted mass density of the base materials, in particular iodine;

    - The noise metric evaluates Incept-net's ability to reduce image noise by comparing its performance with other methods;

    - Total mass density conservation evaluates Incept-net's ability to conserve total mass density in images;

    - Parameter efficiency compares the number of parameters used by Incept-net with other methods, such as U-net. A smaller number of parameters with better performance indicates greater efficiency and generalization capacity of the proposed CNN.\\

    \hline
    Zhou et al. (2019) \cite{Zhou2019_artifacts} & 20 patients  & Investigate the effects of employing a tin filter in conjunction with high-energy-threshold images from a PCCT system & Artifact percentage, noise levels, visualization scores by radiologists, MAE and nominal mass densities & - Artifact percentage measures the proportion of metal artifacts present in the CT images. A lower artifact percentage indicates a reduction in undesirable artifacts, such as streaks and dark/bright regions caused by metal implants;
    
    - Noise Levels refers to random variations in pixel intensity and can impact the clarity of CT images. Lower noise levels are generally desirable as they contribute to clearer and more diagnostically valuable images;

    - Radiologists subjectively evaluate the visualization of specific anatomical structures, including the cortex, trabeculae, and implant-trabecular interface;

    - MAE quantifies the average absolute difference between the predicted and actual mass densities of basis materials. A lower MAE indicates greater accuracy in material density estimation, reflecting the effectiveness of the CNN in providing reliable density information.\\
    \hline
    Anhaus et al. (2022) \cite{Anhaus2022} & Phantom & Assess the optimal reconstruction settings by evaluating the effectiveness of two key approaches: VMI and iMAR & RMSE for artifact quantification, waterfall plots and boxplots, mean HU value and standard deviation, absolute differences calculated for mean and standard deviation & - RMSE measures the average magnitude of errors between predicted and observed values (lower RMSE values indicate better agreement between predicted and observed values);

    - Waterfall plots show the magnitude of different parameters in various categories or conditions, describing the severity or occurrence of artifacts for different imaging conditions, protocols or patient groups;

    - Boxplots provide a visual summary of the distribution of a data set, comparing the variability and central tendency of artifact measurements in different scenarios or groups;

    - The mean HU value and standard deviation are a measure of radiodensity in CT images, and the standard deviation indicates dispersion or variability. They characterize the overall intensity and variability of artefacts in the images;

    - Absolute differences are probably calculated to quantify the dissimilarity or discrepancy between different sets of measurements, to assess how much they vary under different conditions or protocols.\\
    \hline
    Layer et al. (2023) \cite{Layer2023} & 33 scans & Assess the impact of VMI in combination and comparison with iMAR on hip prosthesis‐associated artifacts in PCCT & Qualitative metrics (artifact extent, adjacent soft tissue), and Quantitative metrics (attenuation, standard deviation and adjusted attenuation) & - Artifact extent evaluates the spatial coverage or distribution of metallic artifacts around the hip prosthesis in the reconstructed images;

    - Adjacent soft tissue assessment evaluates the impact of artifacts on nearby soft tissues, providing information on potential distortion;

    - Attenuation values indicate the degree of attenuation of the X-ray beam in different structures;

    - Standard deviation values provide information on how noise levels vary between the different reconstructed images, helping to assess the overall quality of the image;

    - Adjusted attenuation reflects the difference in attenuation between the tissue with artifacts and the corresponding tissue without artifacts.\\
    \hline
    Björkman et al. (2023) \cite{Björkman} & A bovine knee with a stainless steel plate and screws & Determine the optimal imaging parameters for a PCCT and to compare it with an EICT in terms of image quality and severity of metal artifacts & Qualitative metrics (visualization of structures and artifact severity) and Quantitative metrics (Hounsfield Unit value variation, statistical tests (Friedman test, Wilcoxon signed-rank test) and image quality scores) & - Radiologists assess the clarity and detail with which structures were visualized in the images produced and rate the severity of metal artifacts in the images in both systems;

    - HU value variation within ROIs in artifact-affected areas provides a numerical measure of the variability in HU values, offering insights into the image quality and potential artifacts;

    - Statistical tests (Friedman test, Wilcoxon Signed-Rank Test) assess the significance of differences between imaging conditions;

    - Image quality scores represent a numerical scale reflecting the perceived image quality of both systems.\\
    \hline
    Patzer et al. (2023) \cite{Patzer2023} & 50 patients & Evaluate the effectiveness of VMI, iMAR, and their combinations in PCCT scans of patients with dental implants & Attenuation and noise measurements, subjective evaluation, and quantitative measurements of HU for artifact reduction and denoising & - Attenuation and noise measurements evaluates how well the imaging techniques reduce artifacts related to dental implants;

    - Three readers subjectively assess artifact extent and soft tissue interpretability, providing a real-world assessment of the clinical relevance and usability of the imaging techniques;

    - Quantitative Measurements of HU for Artifact Reduction and Denoising provide specific numerical data on how well VMI and iMAR contribute to reducing artifacts and improving image quality.\\
\end{longtable}

\newpage
\subsection{Material Decomposition}
\paragraph{}
PCCT imaging is generally associated with a better breakdown of materials compared to traditional EICT techniques. The main advantage of PCCT lies in its ability to capture detailed information about the energy spectrum of X-ray photons, which leads to better discrimination and decomposition of materials.  Other properties of PCCT allow for a better understanding of the breakdown of materials and tissues, such as improved contrast, which makes it easier to distinguish structures with similar X-ray attenuation values in conventional CT; mono-energetic virtual imaging allows for the generation of monochrome virtual images at specific energy levels, providing greater visibility of certain materials and improving diagnostic accuracy. The ability to analyze the energy spectrum helps reduce artifacts, particularly metallic artifacts, leading to clearer and more accurate images.
\subparagraph{}
Material decomposition algorithms have been developed to provide additional information on tissue composition and make more informed diagnostic decisions while studying PCCT. These algorithms use advanced techniques, such as deep convolutional neural networks and segmentation algorithms, to accurately classify different groups of materials, including breast microcalcifications \cite{Landsmann2022_decomposition_breast}. During phantom studies, PCCT demonstrated precise measurements of material concentrations for iodine, gadolinium, and calcium \cite{Curtis2019}.
\subparagraph{}
Wang et al. presents a thermometry algorithm for PCCT that can accurately predict the temperatures of unknown materials. This is done based on the spectrally resolved linear attenuation coefficients of a set of known base materials at various temperatures. This result is particularly relevant to surgery, as it offers a solution for dealing with the variability of tissue properties without the need to calibrate the tissue in the body in real-time \cite{Wang2023}.
\subparagraph{}
In order to improve a limitation of PCCT related to spectral distortions that affect the accuracy of material decomposition, the article by Nadkarni et al. proposes and demonstrates a solution that uses a DL approach to compensate for these spectral distortions and improve the accuracy of material decomposition in PCCT: Iter2Decomp.
This algorithm showed the best results compared to other methods, including matrix inversion decomposition \cite{Nadkarni2022_decomposition}.
\subparagraph{}
Wu et al. conducted a study focusing on multi-material decomposition in spectral  PCCT using a DL approach. The researchers proposed a method employing improved Fully Convolutional DenseNets to categorize and measure various materials. Their experimental results demonstrated that this approach outperformed traditional methods, particularly in high noise levels, effectively identifying bone, lung, and soft tissue compared to the basis material decomposition based on post-reconstruction space. The findings suggest that the proposed deep learning-based approach holds promise for improving spectral CT material decomposition and may contribute to establishing guidelines for future multi-material decomposition methods in this imaging context \cite{Wu2019}.
\subparagraph{}
Bhattarai et al. explored the diagnostic potential of advanced PCCT for the assessment of osteoporosis. Traditional methods, such as dual X-ray absorptiometry, focus on limited parameters, whereas PCCT, with its high spectral sensitivity, provides qualitative and quantitative information on bone microarchitecture and composition. The study demonstrated the accuracy of PCCT in breaking down calcium content in hydroxyapatite phantoms and rats. It revealed substantial distinctions in morphological parameters between untreated and treated tibial bones. It demonstrated the efficacy of PCCT in both qualitative and quantitative assessment of bone conditions, enabling three-dimensional visualization of the distribution of calcium and water components in bone. PCCT offers an improved way to diagnose bone disease, monitor treatment progress and assess fracture risk \cite{Bhattarai2023}.

\begin{longtable}{|p{2cm}|p{1,6cm}|p{2.2cm}|p{3.2cm}|p{5.3cm}|}
\captionsetup{singlelinecheck=false, justification=raggedright}
\caption{Studies and respective clinical outcomes regarding the material decomposition application of PCCT.}\\
    \hline
    \multicolumn{1}{|c|}{\textbf{Study}} &
    \multicolumn{1}{c|}{\textbf{Scans}} &
    \multicolumn{1}{c|}{\textbf{Purpose}} &
    \multicolumn{1}{c|}{\textbf{Metrics}} &
    \multicolumn{1}{c|}{\textbf{Metric Explanation}}\\
    \hline
    \endfirsthead
    \hline \multicolumn{5}{r}{{(Continues on next page)}} \\ 
    \endfoot
    \hline
    \endlastfoot
    Landsmann et al. (2022) \cite{Landsmann2022_decomposition_breast} & 634 examinations of 317 patients & Investigate the potential of a ML algorithm to classify parenchymal density in breast-PCCT, using a deep CNN & Quantitative metrics (accuracy, ICC, Kappa values, AUC) and qualitative metrics (four-level density scale) to evaluate the performance of the dCNN & - Quantitative metrics: accuracy evaluate the overall performance of the dCNN, ICC and Kappa values assess the agreement between the dCNN and human readers in classifying parenchymal density (ICC more suitable for continuous data, while kappa is more applicable to categorical data), AUC evaluate the diagnostic performance of both human readers and the dCNN;
    
    - Qualitative metrics: a categorization scale ranging from A to D, representing different levels of parenchymal density in order to develop a dCNN model for each category. \\
    
    \hline
    Curtis et al. (2019) \cite{Curtis2019} & Phantom & Demonstrate the ability of PCCT to differentiate multiple mixed compositions, specifically contrast agents (gadolinium and iodine) and tissue components (calcium and water) & Correlation metric, AUC, Root-Mean-Squared Error (RMSE), Visual Assessment - Color Material Concentration Maps & - Correlation metric assess the strength of correlation between the measured x-ray attenuation in each photon energy bin and known concentrations in the calibration phantom during multiple linear regression;
    
    - AUC quantify the accuracy of material identification (AUC greater than 0.95 indicates highly accurate material identification);
    
    - RMSE measure the accuracy of quantitative estimates of material concentrations in mixed compositions;
    
    - Visual assessment allows for qualitative evaluation and differentiation of each material in the sample phantom.\\
    \hline
    Wang et al. (2023) \cite{Wang2023} & Phantom & Explore a new approach to real-time 3D temperature visualization during surgical procedures, using a neural network to predict temperature based on thermal features & Mean Absolute Error (MAE) on validation data and on testing set & - MAE values indicate the average magnitude of errors in temperature predictions on both the validation and testing datasets (lower MAE values are indicative of better predictive performance). \\
    \hline
    Nadkarni et al. (2022) \cite{Nadkarni2022_decomposition} & Model of soft tissue sarcoma that resembles human undifferentiated pleomorphic sarcoma & Propose a DL approach to compensate for the spectral distortions of PCCT and increase the accuracy of material decomposition & RMSE, SSIM, CNR, Modulation Transfer Function (MTF), Qualitative visualization and comparison with current standard & - RMSE measures the average of the squared differences between predicted and actual values;
    
    - SSIM evaluates the quality of generated images compared to the reference image;
    
    - CNR assess the ability to distinguish different materials;
    
    - MTF measures the imaging system's ability to reproduce high-frequency details;
    
    - Qualitative visualization shows the quality of the predictions;

    - The model's performance is compared with the conventional approach of decomposition using matrix inversion on the PCCT. \\
    \hline
    Wu et al. (2019) \cite{Wu2019} & mouse & Proposes a novel approach based on an improved Fully Convolutional DenseNets (FC-DenseNets) for classifying and quantifying different materials, such as bone, lung, and soft tissue & Pixel Accuracy (PA) and Intersection over Union (IoU) or Jaccard Index & - Pa measures the percentage of correctly classified pixels in the image. It provides a general indication of how well the model performs in terms of pixel-wise classification accuracy;

    - IoU measures the overlap between the predicted segmentation mask and the ground truth. It is calculated as the intersection of the predicted and true positive regions divided by the union of these regions. \\
    \hline
    Bhattarai et al. (2023) \cite{Bhattarai2023} & 32 rats & Investigate the diagnostic usefulness of advanced PCCT for assessing bone diseases, in particular osteoporosis & MicroCT metrics: morphological parameters (bone percentage, total porosity, trabecular thickness (Tb.Th), trabecular separation (Tb.Sp), and bone mineral density (BMD)). PCCT metrics: the calcium and water content in the bone. Correlation metrics, statistical analysis (One-way ANOVA, Shapiro–Wilk tests) & - Morphological parameters reveal structural details of bone, highlighting significant distinctions between untreated and ovariectomized samples;
    
    - PCCT metrics and correlation metrics provide insights into bone tissue composition, showing strong correlations that affirm PCCT reliability in characterizing bone composition and its association with bone mineral density;
    
    - One-way ANOVA compares statistical differences between untreated and ovariectomized groups, while Shapiro-Wilk tests confirm the normal distribution of scan data, ensuring the validity of subsequent statistical analyses. \\
\end{longtable}

\newpage
\section{The future of the PCCT}
\paragraph{}
The trajectory of PCCT is marked by both promise and challenges, presenting an attractive scenario for future developments. 
One notable challenge lies in the limited availability of clinical and pre-clinical data, which makes it difficult to progress specific bioengineering projects. The three-dimensional nature of CT data contrasts with the predominant use of two-dimensional processing in most DL projects due to computational constraints. In addition, ethical and regulatory considerations pose ongoing challenges in data acquisition and utilisation. However, advances in hardware and algorithms offer a promising avenue to overcome these limitations, promoting the evolution of new and improved applications \cite{Clark2021}.
\subparagraph{}
The development of a PCD capable of managing the high count rates required for rapid CT acquisitions is a formidable challenge. Pulse stacking, a phenomenon in which several X-rays are recorded as a single high-energy photon due to rapid detection, poses a threat to accurate spectrum detection. Non-ideal detector responses, including the stochastic generation of electron/hole pairs and charge sharing between adjacent elements, further complicate accurate spectrum detection. Spatial defocusing and the detector's low energy resolution add to the complexities. Despite these obstacles, ongoing research aims to find solutions with the potential for improved detector performance and better spectral separation \cite{McCollough2023}.
The concept of "virtual biopsy" is broadening the scope of PCCT applications. PCCT's ability to provide minute details could revolutionise the diagnostic process by enabling non-invasive, high-resolution examinations. This concept aligns with the evolving landscape of personalised medicine, where personalised knowledge contributes to more accurate diagnoses and treatment plans.
When contemplating the future of PCCT, there is a call for the standardisation of research. Understanding the various metrics and outcomes measured in existing articles is crucial to establishing common ground in the scientific community. Standardisation could facilitate collaboration, comparability and a more coherent advancement of PCCT, ensuring that results are universally applicable and contributing to the accumulated knowledge in this field.
\subparagraph{}
Looking to the future, the symbiosis between PCCT and DL promises interesting prospects. The integration of PCCT and DL technologies could give rise to innovative applications in the field of imaging. This synergy could lead to the development of personalised imaging protocols, real-time decision support systems and advanced material decomposition techniques. These advances have the potential to transform the medical imaging landscape, offering new solutions to meet today's challenges.

\section{Conclusion}
\paragraph{}
PCCT demonstrates its effectiveness as a cutting-edge tool in pre-clinical studies, providing a comprehensive view of anatomy, and functionality details. The integration of advanced computational techniques, in particular DL, increases the reliability of PCCT, allowing for more precise control of radiation exposure in medical imaging. Although the superior image quality of PCCT does not necessarily extract more radiomic features, it significantly improves their detectability and clarity, enhancing the quality of radiomic analysis.
The complexity of PCCT data requires algorithms adapted to specific situations. The synergy between DL and PCCT has a positive impact on image quality through advanced reconstruction and noise reduction, and is particularly beneficial in decomposing material for cases such as calcification and osteoporosis. However, the literature on PCCT remains limited, due to recent validation, scarcity of data, computational requirements and regulatory and ethical considerations. Overcoming these challenges is key to unlocking the transformative capabilities of this integrated technology.

\newpage
\bibliographystyle{unsrt}  
\bibliography{references}

\end{document}